\documentclass[aps,pra,showpacs,groupedaddress,floatfix,nofootinbib]{revtex4}

\usepackage{graphicx}

\usepackage{dcolumn}

\begin{document}

\title{On the eigenvalues of some non-Hermitian Hamiltonians with space-time
symmetry}
\author{Paolo Amore}

\email{paolo.amore@gmail.com}

\affiliation{Facultad de Ciencias, Universidad de Colima, Bernal
D\'{i}az del Castillo 340, Colima, Colima, Mexico.}

\author{Francisco M. Fern\'andez}

\email[Corresponding author: ]{fernande@quimica.unlp.edu.ar}

\author{Javier Garcia}

\email{jgarcia@inifta.unlp.edu.ar}

\affiliation{INIFTA (UNLP, CCT La Plata-CONICET), Divisi\'on
Qu\'imica Te\'orica, Blvd. 113 y 64 S/N, Sucursal 4, Casilla de
Correo 16, 1900 La Plata, Argentina}

\begin{abstract}
We calculate the eigenvalues of some two-dimensional non-Hermitian
Hamiltonians by means of a pseudospectral method and
straightforward diagonalization of the Hamiltonian matrix in a
suitable basis set. Both sets of results agree remarkably well but
differ considerably from the eigenvalues obtained some time ago by
other authors. In particular, we do not observe the multiple phase
transitions claimed to occur in one of the anharmonic oscillators.
\end{abstract}


\maketitle

\section{Introduction}

\label{sec:intro}

In a recent paper, Klaiman and Cederbaun\cite{KC08} studied the spectrum of
non-Hermitian Hamiltonians $H=H_{0}+i\lambda W$ by means of the point-group
symmetries of the Hermitian $H_{0}$ and non-Hermitian $W$ parts. They showed
that, in principle, the symmetry properties of the Hamiltonian are
responsible for the appearance of real eigenvalues in the spectrum of the
non-Hermitian Hamiltonian $H$. To this end they constructed an effective
energy-dependent Hermitian Hamiltonian that exhibits the same real spectrum
as the non-Hermitian one. They illustrated the main theoretical results by
means of suitable chosen examples. One of them is of great interest because
it exhibits multiple phase transitions. As the parameter $\lambda $
increases two real eigenvalues approach each other, coalesce and become a
pair of complex conjugate numbers. That is to say, the space-time symmetry
is broken beyond the coalescence point. However, on increasing $\lambda $
those same complex eigenvalues become real again, separate, just to approach
each other again and coalesce at a larger value of $\lambda $.

The purpose of this paper is a critical discussion of those results. In Sec.~%
\ref{sec:models} we outline the point-group symmetry of the models
considered by Klaiman and Cederbaum\cite{KC08}. In Sec.~\ref{sec:results} we
compare present results with the ones of those authors and briefly discuss
an example not considered by them. In Sec.~\ref{sec:conclusions} we draw
conclusions.

\section{The models}

\label{sec:models}

Three of the examples considered by Klaiman and Cederbaun\cite{KC08} are
based on the Hamiltonian
\begin{eqnarray}
H &=&H_{0}+i\lambda W,  \nonumber \\
H_{0} &=&-\frac{1}{2}\left( \partial _{x}^{2}+\partial _{y}^{2}\right)
+\alpha _{x}x^{4}+\alpha _{y}y^{4},  \label{eq:H}
\end{eqnarray}
where $\alpha _{x}=1$ and $\alpha _{y}=\sqrt{2}$. They wrote the
eigenvectors of $H_{0}$ formally as
\begin{equation}
\left| n_{x},n_{y}\right\rangle =\left| n_{x}\right\rangle \otimes \left|
n_{y}\right\rangle ,  \label{eq:|nx,ny>}
\end{equation}
where $n_{x},n_{y}=0,1,\ldots $ and $\left| n_{x}\right\rangle $, $\left|
n_{y}\right\rangle $ denote the eigenvectors of the $x$- and $y$-quartic
oscillators, respectively.

They described the symmetry of $H_{0}$ by means of the point group $D_{2h}$
(isomorphic to $C_{2v}$) with symmetry operations $\left\{
E,P,P_{x},P_{y}\right\} $ that are given by the coordinate transformations
\begin{eqnarray}
E &:&\{x,y\}\rightarrow \{x,y\},  \nonumber \\
P &:&\{x,y\}\rightarrow \{-x,-y\},  \nonumber \\
P_{x} &:&\{x,y\}\rightarrow \{-x,y\},  \nonumber \\
P_{y} &:&\{x,y\}\rightarrow \{x,-y\}.  \label{eq:sym_op}
\end{eqnarray}
It follows from
\begin{eqnarray}
E\left| n_{x},n_{y}\right\rangle &=&\left| n_{x},n_{y}\right\rangle ,
\nonumber \\
P\left| n_{x},n_{y}\right\rangle &=&(-1)^{n_{x}+n_{y}}\left|
n_{x},n_{y}\right\rangle ,  \nonumber \\
P_{x}\left| n_{x},n_{y}\right\rangle &=&(-1)^{n_{x}}\left|
n_{x},n_{y}\right\rangle ,  \nonumber \\
P_{y}\left| n_{x},n_{y}\right\rangle &=&(-1)^{n_{y}}\left|
n_{x},n_{y}\right\rangle ,  \label{eq:O|nx,ny>}
\end{eqnarray}
that the eigenvectors are bases for the irreducible representations $A_{g}$,
$B_{g}$, $A_{u}$ or $B_{u}$ when $(n_{x},n_{y})$ is (even, even), (odd,
odd), (even, odd) or (odd, even), respectively.

\section{Results}

\label{sec:results}

Before discussing the non-Hermitian Hamiltonians considered by Klaiman and
Cederbaun\cite{KC08} we first focus on the Hermitian Hamiltonian $H_{0}$.
Since $\alpha _{y}>\alpha _{x}$ it is clear that $E_{10}^{(0)}$ ($B_{u}$)$%
<E_{01}^{(0)}$ ($A_{u}$). Surprisingly their figures 3 and 4 show exactly
the reverse order. Besides, the same level order appears in Fig. 5 where the
authors labelled the eigenvalues by means of the point group $C_{i}$ instead
of $D_{2h}$.

We calculated the lowest eigenvalues $E_{n_{x}n_{y}}^{(0)}$ of $H_{0}$ by
three completely different approaches: the Riccati-Pad\'{e} method (RPM)\cite
{FMT89a, FMT89b}, a pseudospectral method\cite{AF10} and the straightforward
diagonalization method (DM) using a basis set of products of eigenfunctions
of the harmonic oscillator $H_{HO}=p^{2}+q^{2}$. The three methods agree
remarkably well for $\lambda =0$ and the latter two ones for all $\lambda $
(the RPM does not apply to nonseparable problems).

Table~\ref{tab:Emn(0)} shows the lowest eigenvalues of $H_{0}$ as well as
the symmetry of the corresponding eigenfunctions according to the point
groups $C_{i}$ and $D_{2h}$. By simple inspection it is clear that the
results of this table do not agree with those for $\lambda =0$ in figures 3,
4, and 5 of Ref.\cite{KC08} in agreement with the discussion above.

We first consider the non-Hermitian perturbation $W=xy$ that is invariant
with respect to $P$: $PWP=W$. On the other hand, the whole Hamiltonian (\ref
{eq:H}) is invariant under two antiunitary transformations $A_{x}=TP_{x}$
and $A_{y}=TP_{y}$, where $T$ is the time-reversal operator. According to
the authors it exhibits two space-time symmetries that are a generalization
of the well known PT symmetry\cite{KC08}. In this case the states that
transform as $A_{g}$ ($A_{u}$) couple to states that transform as $B_{g}$ ($%
B_{u}$). The authors illustrate such couplings in their Fig. 3 but, as
discussed above, some of the labels of the lines $E_{mn}(\lambda )$ appear
to exhibit a reverse order and the numerical values of the eigenvalues $%
E_{mn}(0)$ do not appear to agree with present calculation displayed in
Table~\ref{tab:Emn(0)}.

The second example is given by $W=x^{2}y$. In this case the states $A_{g}$ ($%
B_{g}$) couple with the $A_{u}$ ($B_{u}$) ones as shown in Fig. 4 in the
paper by Klaiman and Cederbaum\cite{KC08}. The eigenvalues of $H_{0}$
exhibit the discrepancy already discussed above.

The non-Hermitian perturbation $W=x^{2}y+xy^{2}$ is of special interest
because the authors identified pairs of states that are real for $0<\lambda
<\lambda _{b}$, coalesce at $\lambda _{b}$ and become complex conjugate for $%
\lambda _{b}<\lambda <\lambda _{c}$, then real again for $\lambda
_{c}<\lambda <\lambda _{f}$ and coalesce again at $\lambda =\lambda _{f}$ to
become complex conjugate once more. Bender et al\cite{BGOPY13} have recently
discussed such consecutive phase transitions in the case of classical and
quantum-mechanical linearly-coupled harmonic oscillators (see also \cite{F14}%
). We calculated the same eigenvalues $E_{mn}(\lambda )$ in the same range
of values of $\lambda $ and did not find any of the multiple phase
transitions mentioned by Klaiman and Cederbaum. Fig.~\ref{fig:fig5} shows
present results that exhibit the customary phase transitions for
multidimensional oscillators\cite{BW12}.

Finally, we want to discuss a problem that was not considered by Klaiman and
Cederbaum\cite{KC08} but may be of interest. When $\alpha _{x}=\alpha _{y}=1$
the Hamiltonian $H_{0}$ is invariant under the unitary transformations of
the point group $C_{4v}$. This group exhibits a degenerate irreducible
representation $E$ and, therefore, is beyond the discussion of the paper of
Klaiman and Cederbaum\cite{KC08}. However, we deem it worth mentioning it
here as another example of those discussed by Fern\'{a}ndez and Garcia\cite
{FG14,FG13}. In this case the non-Hermitian perturbation $W=xy$ (with point
group $C_{2v}$) couples the degenerate eigenvectors $\left|
2m,2n+1\right\rangle $ and $\left| 2m+1,2n\right\rangle $ and the $ST$%
-symmetric non-Hermitian operator (\ref{eq:H}) exhibits complex eigenvalues
for all $\lambda >0$. More precisely, some of the eigenvectors of $H_{0}$
belonging to the irreducible representation $E$ with real eigenvalues are
coupled by the non-Hermitian perturbation and become eigenvectors of $H$
belonging to the irreducible representations $B_{1}$ and $B_{2}$ with
complex eigenvalues. As argued by Fern\'{a}ndez and Garcia the $ST$ symmetry
is not as robust as the $PT$ one (were $P$ is the inversion operation in the
point group).

\section{Conclusions}

\label{sec:conclusions}

\begin{figure}[]
\begin{center}
\includegraphics[width=6cm]{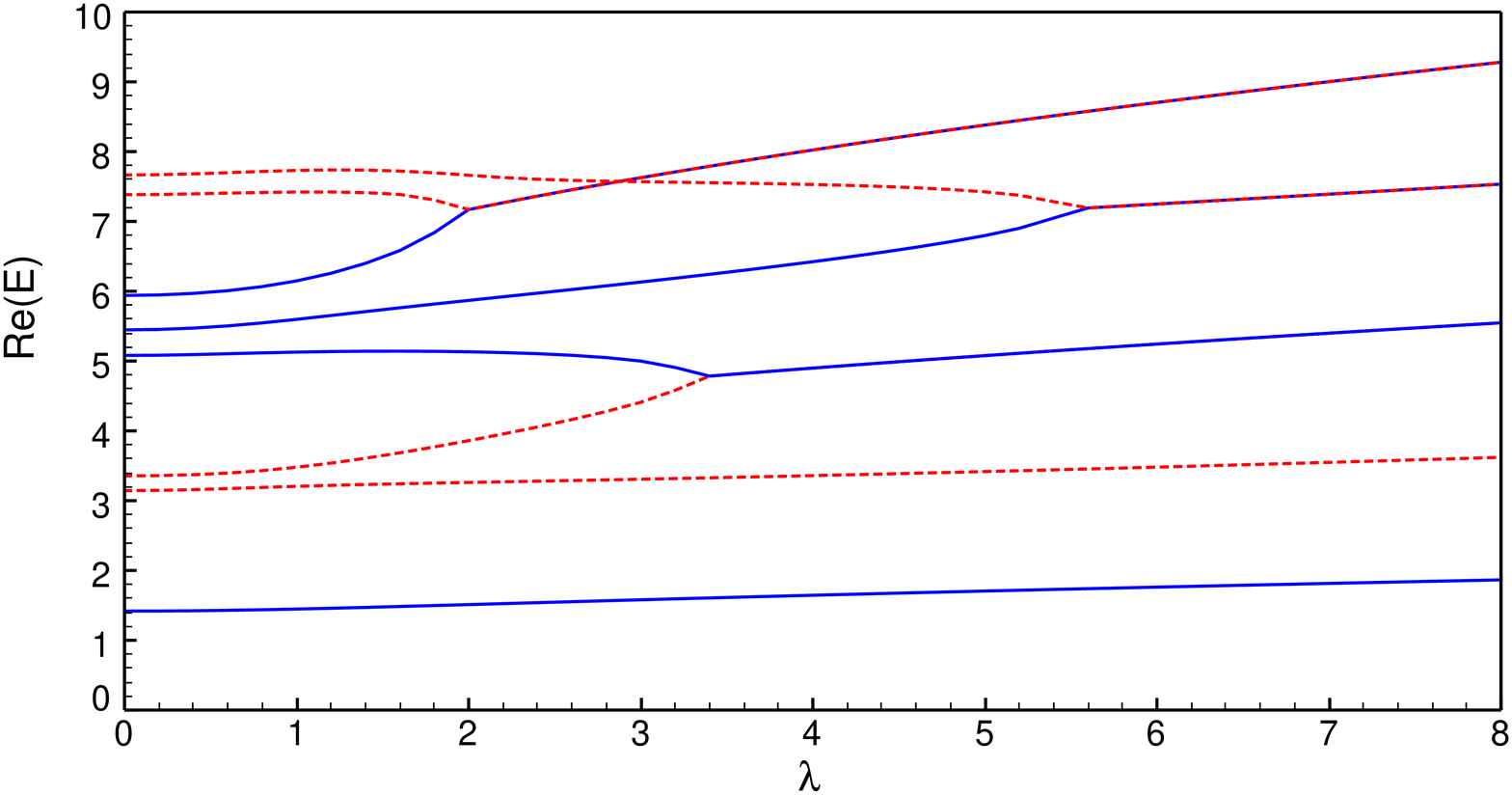}
\includegraphics[width=6cm]{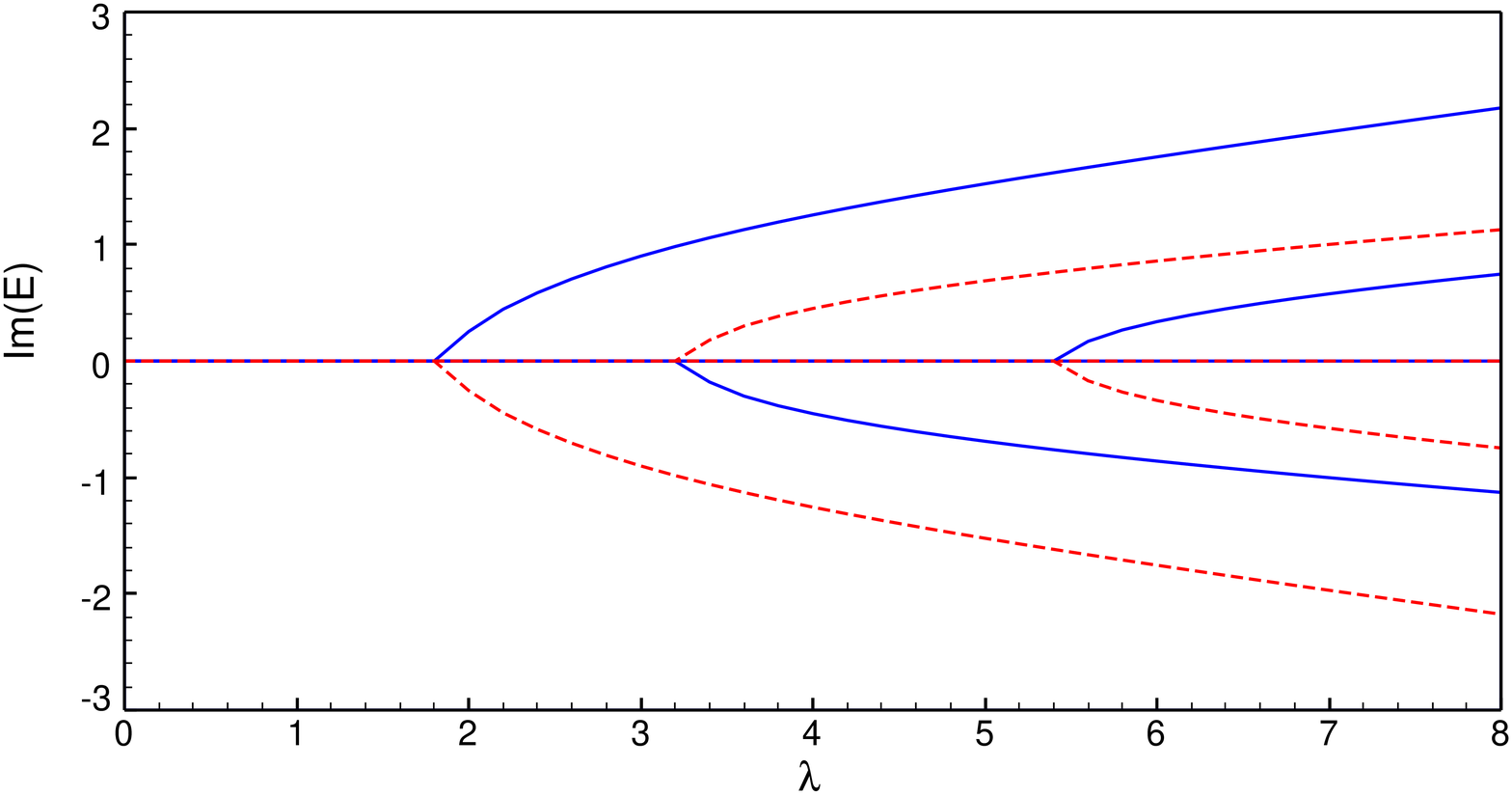}
\end{center}
\caption{First eight eigenvalues of the non-Hermitian Hamiltonian (\ref{eq:H}%
) with $W=x y^2+x^2 y$. The continuous blue lines and dashed red ones
indicate states with symmetry $A_g$ and $A_u$, respectively.}
\label{fig:fig5}
\end{figure}

In this paper we carried out three completely different calculations of the
eigenvalues and eigenfunctions of the Hermitian operator $H_{0}$ and two of
them for the eigenvalues and eigenfunctions of the non-Hermitian operator (%
\ref{eq:H}) with three non-Hermitian perturbations $W$. The agreement of the
results provided by those methods makes us confident of their accuracy.
Present results do not agree with those of Klaiman and Cederbaum\cite{KC08}.
Straightforward comparison of the results in Table~\ref{tab:Emn(0)} with
those in figures 3, 4 and 5 of Ref.\cite{KC08} shows that the magnitude of
the eigenvalues and the level ordering are quite different. Present
eigenvalues $E_{mn}(\lambda )$ for $W=x^{2}y+xy^{2}$ displayed in Fig.~\ref
{fig:fig5} do not exhibit the multiple phase transitions discussed by those
authors but the well-known symmetry breaking at exceptional points common to
other two-dimensional PT-symmetric oscillators\cite{BW12}.

In addition to all that, we have also shown that the $ST$ symmetry proposed
by Klaiman and Cederbaum\cite{KC08} is not as robust as the $PT$ one\cite
{BW12}. In two recent papers Fern\'{a}ndez and Garcia\cite{FG14,FG13} have
already discussed two other $ST$-symmetric cases that exhibit phase
transitions at the trivial Hermitian limit. Those authors also argued that
the coupling of the degenerate states of $H_{0}$ to produce complex
eigenvalues will not take place when $PWP=-W$.

\begin{table}[]
\caption{Lowest eigenvalues $E^{(0)}_{n_xn_y}$ of $H_0$ and the symmetry of
the eigenvectors according to the point groups $C_i$ and $D_{2h}$.}
\label{tab:Emn(0)}%
\begin{tabular}{llD{.}{.}{25}ll}
\hline
  \multicolumn{1}{c}{$n_x$} & \multicolumn{1}{c}{$n_y$}& \multicolumn{1}{c}{$E^{(0)}_{n_xn_y}$}
  & \multicolumn{1}{c}{$C_i$}&\multicolumn{1}{c}{$D_{2h}$}\\

\hline

0&  0 &   1.4177754838502863327 &  Ag &  Ag  \\
1&  0 &   3.1434332411768123405 &  Au &  Bu  \\
0&  1 &   3.3547608248199776054 &  Au &  Au  \\
1&  1 &   5.0804185821465036132 &  Ag &  Bg  \\
2&  0 &   5.4465846115581554207 &  Ag &  Ag  \\
0&  2 &   5.9399608295847593064 &  Ag &  Ag  \\
2&  1 &   7.3835699525278466935 &  Au &  Au  \\
1&  2 &   7.6656185869112853142 &  Au &  Bu  \\
3&  0 &   8.0855192199216020234 &  Au &  Bu  \\
0&  3 &   8.902064775442886576  &  Au &  Au  \\
2&  2 &   9.9687699572926283944 &  Ag &  Ag  \\
3&  1 &   10.022504560891293296 &  Ag &  Bg  \\
1&  3 &   10.627722532769412583 &  Ag &  Bg  \\
4&  0 &   10.9940976801332803   &  Ag &  Ag  \\
0&  4 &   12.166833711560609078 &  Ag &  Ag  \\
3&  2 &   12.607704565656074997 &  Au &  Bu  \\
2&  3 &   12.930873903150755664 &  Au &  Au  \\
4&  1 &   12.931083021102971573 &  Au &  Au  \\
1&  4 &   13.892491468887135086 &  Au &  Bu  \\
5&  0 &   14.12912577729584261  &  Au &  Bu  \\
4&  2 &   15.516283025867753274 &  Ag &  Ag  \\
3&  3 &   15.569808511514202266 &  Ag &  Bg  \\
0&  5 &   15.685783771009134747 &  Au &  Au  \\

\end{tabular}
\end{table}

\end{document}